\documentclass{elsarticle}\usepackage[]{graphicx}\usepackage[]{xcolor}
\makeatletter
\def\maxwidth{ %
  \ifdim\Gin@nat@width>\linewidth
    \linewidth
  \else
    \Gin@nat@width
  \fi
}
\makeatother

\definecolor{fgcolor}{rgb}{0.345, 0.345, 0.345}

\usepackage{framed}
\makeatletter
\newenvironment{kframe}{%
 \def\at@end@of@kframe{}%
 \ifinner\ifhmode%
  \def\at@end@of@kframe{\end{minipage}}%
  \begin{minipage}{\columnwidth}%
 \fi\fi%
 \def\FrameCommand##1{\hskip\@totalleftmargin \hskip-\fboxsep
 \colorbox{shadecolor}{##1}\hskip-\fboxsep
     \hskip-\linewidth \hskip-\@totalleftmargin \hskip\columnwidth}%
 \MakeFramed {\advance\hsize-\width
   \@totalleftmargin\z@ \linewidth\hsize
   \@setminipage}}%
 {\par\unskip\endMakeFramed%
 \at@end@of@kframe}
\makeatother

\definecolor{shadecolor}{rgb}{.97, .97, .97}
\definecolor{messagecolor}{rgb}{0, 0, 0}
\definecolor{warningcolor}{rgb}{1, 0, 1}
\definecolor{errorcolor}{rgb}{1, 0, 0}
\newenvironment{knitrout}{}{} 

\usepackage{alltt}
\usepackage{amsmath,amsthm,amssymb}
\usepackage{booktabs}
\usepackage{array}
\usepackage{multirow}
\usepackage{placeins}
\usepackage{enumitem}
\usepackage{natbib}
\setcitestyle{authoryear}
\usepackage{geometry}
\usepackage{glossaries}
\newgeometry{vmargin={30mm},hmargin={40mm,40mm}}   
\usepackage{subfig}
\usepackage{float}
\usepackage[font=small,skip=0pt]{caption}
\theoremstyle{definition}
\newtheorem{definition}{Definition}

\IfFileExists{upquote.sty}{\usepackage{upquote}}{}
\begin{document}

\title{Classifying Inconsistency in AHP Pairwise Comparison Matrices Using Machine Learning}
\author{Amarnath Bose}
\affiliation{organization={Birla Institute of Management Technology},
addressline={Decision Sciences},
city={Greater Noida}, state={U.P.}, country={India}
}

\begin{abstract}

\noindent
Assessing consistency in Pairwise Comparison Matrices (PCMs) within the Analytical Hierarchy Process (AHP) poses significant challenges when using the traditional Consistency Ratio (CR) method. This study introduces a novel alternative that leverages triadic preference reversals (PR) to provide a more robust and interpretable assessment of consistency. Triadic preference reversals capture inconsistencies between a pair of elements by comparing the direction of preference derived from the global eigenvector with that from a 3×3 submatrix (triad) containing the same pair—highlighting local-global preference conflicts. This method detects a reversal when one eigen ratio exceeds one while another falls below one, signaling inconsistency. We identify two key features: the proportion of preference reversals and the maximum reversal, which mediate the impact of a PCM’s order on its consistency. Using these features, simulated PCMs are clustered into consistent and inconsistent classes through k-means clustering, followed by training a logistic classifier for consistency evaluation. The PR method achieves 97\% accuracy, significantly surpassing the Consistency Ratio (CR) method's 50\%, with a false negative rate of only 2.6\% compared to 5.5\%. These findings demonstrate the PR method's superior accuracy in assessing AHP consistency, thereby enabling more reliable decision-making. The proposed triadic preference reversal (PR) approach is implemented in the R package AHPtools publicly available on the Comprehensive R Archive Network (CRAN). 
\end{abstract}


\maketitle

\section{Introduction}

Evaluating the consistency of Pairwise Comparison Matrices (PCM) in the Analytic Hierarchy Process (AHP) presents complex challenges intertwined with the intricate decision making processes guiding its construction, the scale chosen for pairwise comparisons and the number of alternatives considered. This complicates the selection of consistency metrics impeding the adoption of an objectively indisputable approach for valid consistency classification.

The generation of Pairwise Comparison Matrices (PCMs) in AHP involves purposeful, perception-based assignment of comparison ratios, surpassing the simplicity of randomly generated PCMs. The process of assigning the pairwise comparison ratios demands careful consideration to accurately reflect a decision-makers' subjective judgments and relative preferences for alternative pairs. Decision-makers systematically assess the relative importance of criteria pairs in assigning numerical values that reflect perceived comparative preference for either alternative. Factors like importance of either alternative in achieving a goal, user's preferences and priorities influence the value assigned. 

It is essential to recognise that the pairwise ratios in a PCM reflect a decision-makers' insights and judgment. This intentionality ensures that pairwise comparison matrices represent a decision-maker's perspective, and is therefore the only credible basis for deriving thresholds for consistency. On the contrary, randomly generated PCMs do not capture the nuance of an intentionally generated, logical PCMs. Using random PCMs as a basis for CR consistency assessment naturally results in the derivation of flawed consistency thresholds.

Central to this research is therefore the simulation of `logical PCMs' distinguished from their random counterparts, by using a method discussed in \cite{Bose2022}. Only on the basis of such logical PCMs is it meaningful to address the challenge of valid evaluation of consistency. 

With random PCMs replaced by logical PCMs the stage is set for identifying and operationalizing a valid consistency criteria. This research prioritises and emphasises naturalness and intuitive appeal over the mechanical pedanticity of an eigenvalue-based consistency measure.

As a fundamental measure of consistency we consider reversals in pairwise preferences between elements of the eigenvector of the whole PCM and its order-3 submatrices. 
The intensity of preference reversal, measured by the number of such reversals and the maximum reversal, indicates inconsistency. For an order $n$ PCM the number of order-3 submatrices for a given pair of alternatives is $n{-}2$, which is the number of available third alternatives for any chosen pair. 

Preference reversals considering all the $\binom{n}{2}$ pairs of alternatives serves as an important intuitive indicator of inconsistency; the greater this number for a given order of a logical PCM, the greater is the inconsistency. There are two derivatives from preference reversals for a PCM that inform the consistency of a PCM - a) the proportion of reversals and b) the maximum of the reversals. 

The next challenge is to determine the set of values for the preference reversal metrics that, for any given order, would qualify a PCM to be consistent or otherwise. A machine learning approach based on unsupervised clustering is used in view of the subjectivity of the very concept of consistency. For a training set of PCMs, given the three preference reversal metrics, an unsupervised binary clustering accomplishes the task of assigning the appropriate consistency classification labels. Finally, a binary logistic classification model is developed based on the three preference reversal metrics and the classification labels generated for the training corpus of logical PCMs. The logit model classifies a PCM based on its preference reversal values which can easily be computed. 

The importance of the basis of logical PCMs comes across as critical in this method. In the unsupervised binary clustering we can assume, given the way logical PCMs are constructed, that the PCMs that belong to the inconsistent cluster have some combination of a high number of preference reversals, a high maximum reversal or a high average reversal value, and  can therefore be assumed to be naturally inconsistent. 

\section{Literature Review}

The Analytic Hierarchy Process (AHP), described by Saaty in various works such as \citep{Saaty1977, Saaty2008, Saaty2013}, stands as a widely adopted tool for multi-criteria decision analysis. Structured hierarchies of options, sub-options, and criteria, with preferences expressed through pairwise comparison matrices, facilitate objective decision-making to achieve desired objectives \citep{Wind1980,Vaidya2006,Podvezko2009}. Consistency within the constituent Pairwise Comparison Matrices (PCMs) in an AHP hierarchy is imperative to ensure the resulting priorities accurately reflect decision-makers' preferences \citep{Franek2014,Ishizaka2011}.

The Consistency Ratio (CR) method \citep{Saaty2003,Saaty2008} is most commonly used for consistency evaluation. The CR method involves a sequence of steps, including rescaling the principal eigenvalue, comparing it to the average of simulated random matrices, and applying predetermined threshold values (0.05 for matrices of orders 3 and 4, and 0.10 for higher-order PCMs) to classify consistency. However, using random matrices as benchmarks for consistency classification proves problematic, as they do not accurately represent perception-based, rational choices made by decision-makers. This reliance on arbitrary thresholds raises questions about the robustness of the classification process. Furthermore, the use of the Fundamental Scale for assigning PCM elements presents additional challenges to achieving consistency.

The critical importance of consistency in AHP has spurred considerable interest in its study, resulting in two primary research directions. One involves strategies for enhancing PCM consistency, proposed by \cite{Harker1987,Xu1999,Jarek2016,Kou2014,Ergu2011}. These strategies aim to minimize inconsistency by detecting and adjusting elements contributing most to it. Evolutionary methods such as \cite{Costa2011,Bose2020,Girsang2014} have also been developed to improve the consistency of a PCM.

Another branch of research has focused on critiquing and refining existing consistency measures. \cite{Lane1989,Koczkodaj1993} challenged Saaty's consistency thresholds, while \cite{Vargas2008} highlighted potential errors in associating randomness with inconsistency. \cite{Bozoki2008} critically examined proposed inconsistency metrics, emphasizing the need for further refinement. \cite{Franek2014} explored the impact of judgment scales on consistency, finding variability across different scales. 

Diverse approaches have been proposed to address the limitations of existing consistency evaluation methods. \cite{Dodd1993} proposed statistical acceptance levels, while \cite{Monsuur1997} suggested an intrinsic and scale-independent threshold. Alternative consistency metrics and thresholds have been introduced by \cite{Pelaez2003,Aguaron2003} while \cite{Pelaez2018} developed an intuitive consistency index based on probability distributions. Another approach to consistency proposed by \citet{Bose2023} classified a PCM based on its distance bfrom its benchmark comprising the most consistent, feasible set of comparison ratios representative of the original preferences. In this paper, inconsistency in non-random, `logical' matrices was modeled on the matrix order mediated by the preference range for the matrix alternatives. 

A distinctive line of consistency research (\citet{Szybowski2014, Duszak1994, Koczkodaj2017, Szybowski2016, Csato2019, Aguaron2020}) focuses on triads of alternatives in a pairwise matrix. Koczkodaj \citet{Koczkodaj2017, Duszak1994} proposed an inconsistency measure that directly uses triads of pairwise comparisons instead of eigen-based prioritization. The basic formulation of the inconsistency measure which is elegant and intuitive is
\[
\textit{K(A)} = \max_{i<j<k} \{\min \{ \big|1-\frac{a_{ik}}{a_{ij}a_{jk}}\big|, \big|1-\frac{a_{ij}a_{jk}}{a_{ik}}\big| \}\}
\]
Triad based inconsistency measures are based on evaluation of the intensities of triad-cycles such as ($a_{ij}, a_{jk}, a_{ik}$) with the premise that the inconsistency is more if $a_{ij} a_{jk}$ varies widely from $a_{ik}$.

Given the critical role of consistency in AHP, it is imperative to address the identified issues to achieve a more objective classification approach. The approach should utilize a measure along with a benchmarking scheme that ensures that would ensure a bias free, matrix order agnostic, reasonable consistency classification. We hypothesize that the shortcomings of the CR method stem from its oversimplified eigenvalue-based approach and unrealistic benchmarking with random PCMs. In Section~\ref{methodology}, we propose an approach to overcome these limitations, followed by a demonstration of its effectiveness in Section~\ref{results}.


\section{Methodology}
\label{methodology}



The method of classification proposed draws validity from `logical PCMs' that, unlike random PCMs are similar to what human decision makers would come up with for a matrix of pair-wise comparison ratios. It is quite obvious that logical PCMs would have smaller principal eigenvalues than random PCMs for the same matrix order and would be generally more consistent in a common sense of usage of the term. In order to model consistency in AHP PCMs it is essential to use such logical PCMs, to ensure that any measure indicating consistency would be realistically representative. A simple method of simulating logical PCMs has been described in \cite{Bose2022} and is briefly outlined in Section~\ref{simulationMethod}. 

A more fundamental problem is to characterize consistency of a PCM, and for this purpose we propose a method that uses reversals of pairwise preferences within submatrices of a PCM. This is elaborated in Section~\ref{consistencyCharacterization}.

In the following subsections, we develop these three artefacts. 

\subsection{Intuitive Consistency Characterization for PCMs}
\label{consistencyCharacterization}

In any PCM of order $n$, there are two sets of numeric measures: an input set of $\binom{n}{2}$ comparison ratios, and the derived $n$ evaluated eigen-priorities. Inconsistency can naturally be ascribed to contradictions occurring between the evaluated eigen-priorities. In order to detect such inconsistency we consider PCMs of order three, constructed from all the $\binom{n}{3}$ subset of alternatives from the original $n$. These triplets contain the fundamental indicators of inconsistency for the PCM, occurring when $a_{ij} a_{jk} \neq a_{ik}$. For any tuple (i,j) we consider all $n{-}2$ submatrices of order 3 that contain both alternatives. These submatrices contain both alternatives in the tuple and one of the $n{-}2$ other alternatives. These submatrices are the smallest ones where inconsistencies in eigen-priorities can be detected. For tuple (i,j) $n{-}2$ eigen-domination values can be derived from these order 3 submatrices. A reversal of any of these $n{-}2$ eigen-domination ratios with respect to the original eigen-domination ratio signifies a contradiction, and is a potential indicator of inconsistency. The number of such reversals, ranging from 0 to a maximum of $n{-}2$ is considered as a metric for adjudging consistency.

The principal eigenvector of a PCM suggests the aggregate priorities for alternatives compared in a PCM. For purposes of this article these are referred as `eigen-priorities'. An \textit{alternative $i$} will be said to eigen-dominate \textit{alternative $j$} if $\frac{ep_i}{ep_j}>1$ where $ep_i$ is the eigen-priority for alternative $i$. The choice of $1$ is to ensure a reasonably clear dominance threshold. 

\begin{definition}[\textit{eigen-priority}]
For a PCM of order $n$ this term refers to the values of the principal eigenvector which indicate the relative priorities for the $n$ alternatives.
\end{definition}

\begin{definition}[\textit{eigen-dominance}]
For a PCM of order $n$ an $i^{th}$ alternative will be said to eigen-dominate the $j^{th}$ alternative if $\frac{e_i}{e_j}>1$ where $e_i$ is the eigen-priority for alternative $i$.
\end{definition}

\begin{definition}[\textit{reversal}]
A reversal will be said to occur between two alternatives $i$ and $j$, if one of the two alternatives, say the $i^{th}$, `\textit{eigen-dominates}' the other $j^{th}$ alternative in the entire $n$-element principal eigenvector, whereas for a $n{-}2$ order-3 PCMs involving both these alternatives, the $j^{th}$ alternative eigen-dominates the $i^{th}$ alternative.
\end{definition}

A measure that indicates the severity of inconsistency is the maximum of the $n{-}2$ reversals, computed as $\max\limits_{\substack{1 \leqslant i, j \leqslant n \\ i \neq j}} \frac{e_i}{e_j} \div \frac{e3_i}{e3_j}$, where $e_i$ denotes the eigen-priority computed for the $i^{th}$ alternative in the original order-n PCM, and $e3_i$ denotes the eigen-priority computed for the $i^{th}$ alternative in one of its order 3 submatrices. A larger number indicates a strong reversal in eigen-priority when a subset of three alternatives are chosen from the full $n$ alternatives.

Two indicative measures of consistency for a PCM are used for our classification model, the proportion of reversals and the maximum reversal. 

To find the propensity and extent of eigen-domination for PCMs in order to determine appropriate thresholds for consistency, we need `\textit{natural PCMs}' simulated at scale for each possible order. The method for simulating natural PCMs is outlined in the following section.

\subsection{Developing the Consistency Classification Model}
\label{consistencyModel}
A measure of the degree of consistency for a Pairwise Comparison Matrix is not straight-forward. Moreover, the threshold acceptable for consistency, for any given order of a PCM is subjective.
The more fundamental of the two, the measure of degree of consistency of a PCM is contingent upon factors such as the inter-relationship of the individual matrix elements as well as its overall structure. We have identified two derived indicators as described in Section~\ref{consistencyCharacterization} that indicate the degree of inconsistency of a PCM. 

With these indicators computed for a each of a simulated dataset of 180000 logical PCMs, 20000 for each of orders 4 though 12, we  deploy an unsupervised k-means clustering algorithm to derive the consistency classification. Thereafter a logistic classifier is developed using the generated classification labels.

\subsubsection{Simulating Logical PCMs}

\label{simulationMethod}

For a PCM of order $n$, we generate a random numeric vector, $e$, of size $n$. A PCM generated from this vector, with elements $a_{ij}=\frac{e_i}{e_j}$ is obviously consistent. For each element in this `perfectly consistent PCM' with value greater than 1 we consider its absolute difference with elements of the subset of the Fundamental Scale $\{1,2,3,4,\cdots,8,9\}$. Out of the eight absolute differences thus obtained, a random choice is made from the smallest five. This randomization takes into consideration human errors of judgement; if we had selected only three instead of the five smallest elements, the demands on consistency would have been higher. The element in the PCM is chosen from the subset of the Fundamental Scale, $\{1,2,3,4,\cdots,8,9\}$  corresponding to the random choice made. The transpose of the element is assigned the inverse value.

As an example, we simulate a naturally consistent PCM, $P'$, of order 5. For this, we generate a five element vector, with each element following the random uniform distribution in the range 0 to 1. This is $(0.738, 0.157, 0.828, 0.848, 0.999)$. A consistent PCM, $P$ is constructed from this vector. Thereafter, a logical PCM, $P'$ is generated by following the above procedure for each element of the PCM having a value more than 1.

\begin{align*}
P &= \begin{pmatrix}
1 & 4.71 & 0.89 & 0.87 & 0.74 \\
0.21 & 1 & 0.19 & 0.18 & 0.16 \\
1.12 & 5.28 & 1 & 0.98 & 0.83 \\
1.15 & 5.41 & 1.02 & 1 & 0.85 \\
1.35 & 6.37 & 1.21 & 1.18 & 1 
\end{pmatrix}
& 
P' &= \begin{pmatrix}
1 & 6 & 0.5 & 1 & 0.5 \\
0.17 & 1 & 0.25 & 0.17 & 0.12 \\
2 & 4 & 1 & 0.33 & 1 \\
1 & 6 & 3 & 1 & 1 \\
2 & 8 & 1 & 1 & 1 
\end{pmatrix}
\end{align*}

\subsubsection{Inconsistency Indicators}
\label{inconsistencyindicators}

For pairwise comparison matrices (PCMs) of order four or higher, we use two measures to assess inconsistency. These measures reflect the degree of mismatch between the pairwise priorities in the full matrix and those in its order-3 submatrices. The two measures as mentioned in Section~\ref{consistencyCharacterization} are the proportion of reversals, $prop3Rev$ and the maximum reversal, $max3Rev$.

In the following example the three inconsistency indicators are computed for the following order 5 PCM with the 5 alternatives denoted as $a$, $b$, $c$, $d$ and $e$.

\begin{table}[!ht]
\centering
\begin{tabular}{l|rrrrr|}
& \small{\text{a}} & \small{\text{b}} & \small{\text{c}} & \small{\text{d}} & \small{\text{e}} \\
\hline
\small{\text{a}} & \small{1} & \small{1/5} & \small{1/5} & \small{2} & \small{1/6} \\
\small{\text{b}} & \small{5} & \small{1} & \small{4} & \small{1/3} & \small{4} \\
\small{\text{c}} &\small{5} & \small{1/4} & \small{1} & \small{1} & \small{2} \\
\small{\text{d}} &\small{1/2} & \small{3} & \small{1} & \small{1} & \small{1/9} \\
\small{\text{e}} &\small{6} & \small{1/4} & \small{1/2} & \small{9} & \small{1} \\
\hline
\end{tabular}
\end{table}

The principal eigenvector for this PCM, indicative of the relative preferences, is \\ \small{(0.137, 0.637, 0.335, 0.337, 0.592)}\label{evec}. This shows that, in an aggregate sense, $b$ is the most preferred alternative while $a$ is least preferred.\\

\text{\bf Step 1: submatrices of order 3} 

For the above order-5 PCM, the 10 submatrices of order 3 along with their respective eigenvectors are displayed.

\begin{table}[H]
\begin{tabular}{|p{.25cm}p{.25cm}p{.25cm}|p{.3cm}|p{.25cm}p{.25cm}p{.25cm}|p{.3cm}|p{.25cm}p{.25cm}p{.25cm}|p{.3cm}|p{.25cm}p{.25cm}p{.25cm}|p{.3cm}|p{.25cm}p{.25cm}p{.25cm}|}
& \small{abc} & && & \small{abd} & &&  & \small{abe} & && & \small{acd} & && & \small{ace} & \\

\scriptsize{1} & \scriptsize{1/5} & \scriptsize{1/5} && \scriptsize{1} & \scriptsize{1/5} & \scriptsize{2} && \scriptsize{1} & \scriptsize{1/5} & \scriptsize{1/6} && \scriptsize{1} & \scriptsize{1/5} & \scriptsize{2} &&  \scriptsize{1} & \scriptsize{1/5} & \scriptsize{1/6} \\
\scriptsize{5} & \scriptsize{1} & \scriptsize{4} && \scriptsize{5} & \scriptsize{1} & \scriptsize{1/3} && \scriptsize{5} & \scriptsize{1} & \scriptsize{4} && \scriptsize{5} & \scriptsize{1} & \scriptsize{1} && \scriptsize{5} & \scriptsize{1} & \scriptsize{2} \\
\scriptsize{5} & \scriptsize{1/4} & \scriptsize{1} && \scriptsize{1/2} & \scriptsize{3} & \scriptsize{1}  && \scriptsize{6} & \scriptsize{1/4} & \scriptsize{1} && \scriptsize{1/2} & \scriptsize{1} & \scriptsize{1} && \scriptsize{6} & \scriptsize{1/2} & \scriptsize{1} 
\end{tabular}
\vspace{-0.3cm} 
\begin{tabular}{>{\centering\arraybackslash}m{.25cm}>{\centering\arraybackslash}m{.25cm}>{\centering\arraybackslash}m{.25cm}>{\centering\arraybackslash}m{.35cm}>{\centering\arraybackslash}m{.25cm}>{\centering\arraybackslash}m{.25cm}>{\centering\arraybackslash}m{.25cm}>{\centering\arraybackslash}m{.35cm}>{\centering\arraybackslash}m{.25cm}>{\centering\arraybackslash}m{.25cm}>{\centering\arraybackslash}m{.25cm}>{\centering\arraybackslash}m{.35cm}>{\centering\arraybackslash}m{.25cm}>{\centering\arraybackslash}m{.25cm}>{\centering\arraybackslash}m{.25cm}>{\centering\arraybackslash}m{.35cm}>{\centering\arraybackslash}m{.25cm}>{\centering\arraybackslash}m{.25cm}>{\centering\arraybackslash}m{.25cm}}
 & & && & & && & & && & & && & & \\[-2ex]
\small{a}& \small{b}& \small{c}&& \small{a}& \small{b}& \small{d}&& \small{a} & \small{b} & \small{e} && \small{a}& \small{c}& \small{d}&& \small{a}& \small{c}& \small{e} \\
\scriptsize{.116} & \scriptsize{.923} & \scriptsize{.366} && \scriptsize{.408} & \scriptsize{.657} & \scriptsize{.634} && 
\scriptsize{.109} & \scriptsize{.916} & \scriptsize{.386} && \scriptsize{.364} & \scriptsize{.845} & \scriptsize{.392} && 
\scriptsize{.123} & \scriptsize{.825} & \scriptsize{.552}\\
\cline{1-3} \cline{5-7} \cline{9-11} \cline{13-15} \cline{17-19}
\end{tabular}
\end{table}

\vspace{-0.1cm} 
\begin{table}[H]
\begin{tabular}{|p{.25cm}p{.25cm}p{.25cm}|p{.3cm}|p{.25cm}p{.25cm}p{.25cm}|p{.3cm}|p{.25cm}p{.25cm}p{.25cm}|p{.3cm}|p{.25cm}p{.25cm}p{.25cm}|p{.3cm}|p{.25cm}p{.25cm}p{.25cm}|}
& \small{ade} & && & \small{bcd} & &&  & \small{bce} & && & \small{bde} & && & \small{cde} & \\

\scriptsize{1} & \scriptsize{2} & \scriptsize{1/6} && \scriptsize{1} & \scriptsize{4} & \scriptsize{1/3} && \scriptsize{1} & \scriptsize{4} & \scriptsize{4} && \scriptsize{1} & \scriptsize{1/3} & \scriptsize{4} &&  \scriptsize{1} & \scriptsize{1} & \scriptsize{2} \\
\scriptsize{1/2} & \scriptsize{1} & \scriptsize{1/9} && \scriptsize{1/4} & \scriptsize{1} & \scriptsize{1} && \scriptsize{1/4} & \scriptsize{1} & \scriptsize{2} && \scriptsize{3} & \scriptsize{1} & \scriptsize{1/9} && \scriptsize{1} & \scriptsize{1} & \scriptsize{1/9} \\
\scriptsize{6} & \scriptsize{9} & \scriptsize{1} && \scriptsize{3} & \scriptsize{1} & \scriptsize{1}  && \scriptsize{1/4} & \scriptsize{1/2} & \scriptsize{1} && \scriptsize{1/4} & \scriptsize{9} & \scriptsize{1} && \scriptsize{1/2} & \scriptsize{9} & \scriptsize{1} 
\end{tabular}
\vspace{-0.3cm} 
\begin{tabular}{p{.25cm}p{.25cm}p{.25cm}p{.35cm}p{.25cm}p{.25cm}p{.25cm}p{.35cm}p{.25cm}p{.25cm}p{.25cm}p{.35cm}p{.25cm}p{.25cm}p{.25cm}p{.35cm}p{.25cm}p{.25cm}p{.25cm}}
 & & && & & && & & && & & && & & \\[-2ex]
\small{a}& \small{d}& \small{e}&& \small{b}& \small{c}& \small{d}&& \small{b} & \small{c} & \small{e} && \small{b}& \small{d}& \small{e}&& \small{c}& \small{d}& \small{e} \\
\scriptsize{.180} & \scriptsize{.099} & \scriptsize{.979} && \scriptsize{.573} & \scriptsize{.328} & \scriptsize{.751} && 
\scriptsize{.937} & \scriptsize{.295} & \scriptsize{.186} && \scriptsize{.596} & \scriptsize{.376} & \scriptsize{.710} && 
\scriptsize{.591} & \scriptsize{.226} & \scriptsize{.774} \\
\cline{1-3} \cline{5-7} \cline{9-11} \cline{13-15} \cline{17-19}
\end{tabular}
\end{table} 

In the first of the ten order-3 PCMs for example, the three alternatives, a, b and c are considered. The eigenvector (.116, .923, .366) represents the relative importance of the alternatives $a$, $b$ and $c$.

\vspace{0.5cm}
\text{\bf Step 2: Computing Reversals} \\

For each of the order 3 submatrices from Step 1, we compute reversals, if any.

\begin{table}[H]
\begin{tabular}{|>{\centering\arraybackslash}m{.2cm}>{\centering\arraybackslash}m{1cm}>{\centering\arraybackslash}m{1cm}>{\centering\arraybackslash}m{.5cm}|>{\centering\arraybackslash}m{.07cm}|>{\centering\arraybackslash}m{.2cm}>{\centering\arraybackslash}m{1cm}>{\centering\arraybackslash}m{1cm}>{\centering\arraybackslash}m{.5cm}|>{\centering\arraybackslash}m{.07cm}|>{\centering\arraybackslash}m{.2cm}>{\centering\arraybackslash}m{1cm}>{\centering\arraybackslash}m{1cm}>{\centering\arraybackslash}m{.5cm}|}
\scriptsize{abc} & \scriptsize{full} & \scriptsize{triad} & \scriptsize{rev} && \scriptsize{abd} & \scriptsize{full} & \scriptsize{triad}  & \scriptsize{rev} && \scriptsize{abe} & \scriptsize{full} & \scriptsize{triad}  & \scriptsize{rev} \\
\cline{1-4} \cline{6-9} \cline{11-14}
\scriptsize{ab} & \scriptsize{.137/.637} & \scriptsize{.116/.923} &  \scriptsize{-}   && \scriptsize{ab} & \scriptsize{.137/.637} & \scriptsize{.408/.657} & \scriptsize{-}  && \scriptsize{ab} & \scriptsize{.137/.637} & \scriptsize{.109/.916} & \scriptsize{-} \\
\scriptsize{ac} & \scriptsize{.137/.335} & \scriptsize{.116/.366} & \scriptsize{-}   && \scriptsize{ad} & \scriptsize{.137/.337} & \scriptsize{.408/.634} & \scriptsize{-}  && \scriptsize{ae} & \scriptsize{.137/.592} & \scriptsize{.109/.386} & \scriptsize{-} \\
\scriptsize{bc} & \scriptsize{.637/.335} & \scriptsize{.923/.366} &  \scriptsize{-}  && \scriptsize{bd} & \scriptsize{.637/.337} & \scriptsize{.657/.634} & \scriptsize{-}  && \scriptsize{be} & \scriptsize{.637/.592} & \scriptsize{.916/.386} &\scriptsize{-}  
\end{tabular}
\end{table}

\vspace{-0.3cm} 
\begin{table}[!ht]
\begin{tabular}{|>{\centering\arraybackslash}m{.2cm}>{\centering\arraybackslash}m{1cm}>{\centering\arraybackslash}m{1cm}>{\centering\arraybackslash}m{.5cm}|>{\centering\arraybackslash}m{.07cm}|>{\centering\arraybackslash}m{.2cm}>{\centering\arraybackslash}m{1cm}>{\centering\arraybackslash}m{1cm}>{\centering\arraybackslash}m{.5cm}|>{\centering\arraybackslash}m{.07cm}|>{\centering\arraybackslash}m{.2cm}>{\centering\arraybackslash}m{1cm}>{\centering\arraybackslash}m{1cm}>{\centering\arraybackslash}m{.5cm}|}
\scriptsize{acd} & \scriptsize{full} & \scriptsize{   triad}  & \scriptsize{rev} && \scriptsize{ace} & \scriptsize{full} & \scriptsize{triad}  & \scriptsize{rev} && \scriptsize{ade} & \scriptsize{full} & \scriptsize{triad}  & \scriptsize{rev} \\
\cline{1-4} \cline{6-9} \cline{11-14}
\scriptsize{ac} & \scriptsize{.137/.335} & \scriptsize{.364/.845} &  \scriptsize{-} 
    && \scriptsize{ac} & \scriptsize{.137/.335} & \scriptsize{.123/.825} & \scriptsize{-} 
    && \scriptsize{ad} & \scriptsize{.137/.337} & \scriptsize{.180/.099} & \scriptsize{4.472} \\
    
\scriptsize{ad} & \scriptsize{.137/.337} & \scriptsize{.364/.392} & \scriptsize{-}   && \scriptsize{ae} & \scriptsize{.137/.592} & \scriptsize{.123/.552} & \scriptsize{-}  && \scriptsize{ae} & \scriptsize{.137/.592} & \scriptsize{.180/.979} & \scriptsize{-} \\

\scriptsize{cd} & \scriptsize{.335/.337} & \scriptsize{.845/.392} &  \scriptsize{2.168}  && \scriptsize{ce} & \scriptsize{.335/.592} & \scriptsize{.825/.552} & \scriptsize{2.641}  && \scriptsize{de} & \scriptsize{.337/.592} & \scriptsize{.099/.979} &\scriptsize{-}  
\end{tabular}
\end{table}

\vspace{-0.3cm} 
\begin{table}[!ht]
\begin{tabular}{|>{\centering\arraybackslash}m{.2cm}>{\centering\arraybackslash}m{1cm}>{\centering\arraybackslash}m{1cm}>{\centering\arraybackslash}m{.5cm}|>{\centering\arraybackslash}m{.07cm}|>{\centering\arraybackslash}m{.2cm}>{\centering\arraybackslash}m{1cm}>{\centering\arraybackslash}m{1cm}>{\centering\arraybackslash}m{.5cm}|>{\centering\arraybackslash}m{.07cm}|>{\centering\arraybackslash}m{.2cm}>{\centering\arraybackslash}m{1cm}>{\centering\arraybackslash}m{1cm}>{\centering\arraybackslash}m{.5cm}|}
\scriptsize{bcd} & \scriptsize{full} & \scriptsize{triad}  & \scriptsize{rev} && \scriptsize{bce} & \scriptsize{full} & \scriptsize{triad}  & \scriptsize{rev} && \scriptsize{bde} & \scriptsize{full} & \scriptsize{triad}  & \scriptsize{rev} \\
\cline{1-4} \cline{6-9} \cline{11-14}
\scriptsize{bc} & \scriptsize{.637/.335} & \scriptsize{.573/.328} &  \scriptsize{-}   && \scriptsize{bc} & \scriptsize{.637/.335} & \scriptsize{.937/.295} & \scriptsize{-}  && \scriptsize{bd} & \scriptsize{.637/.337} & \scriptsize{.596/.376} & \scriptsize{-} \\
\scriptsize{bd} & \scriptsize{.637/.337} & \scriptsize{.573/.751} & \scriptsize{2.477}   && \scriptsize{be} & \scriptsize{.637/.592} & \scriptsize{.937/.186} & \scriptsize{-}  && \scriptsize{be} & \scriptsize{.637/.592} & \scriptsize{.596/.710} & \scriptsize{1.282} \\
\scriptsize{cd} & \scriptsize{.335/.337} & \scriptsize{.328/.751} &  \scriptsize{-}  && \scriptsize{ce} & \scriptsize{.335/.592} & \scriptsize{.295/.186} & \scriptsize{2.802}  && \scriptsize{de} & \scriptsize{.337/.592} & \scriptsize{.376/.710} &\scriptsize{-}  
\end{tabular}
\end{table}

\vspace{-0.3cm} 
\begin{table}[!ht]
\begin{tabular}{|>{\centering\arraybackslash}m{.2cm}>{\centering\arraybackslash}m{1cm}>{\centering\arraybackslash}m{1cm}>{\centering\arraybackslash}m{.5cm}|}
\scriptsize{cde} & \scriptsize{full} & \scriptsize{triad}  & \scriptsize{rev}  \\
\cline{1-4}
\scriptsize{cd} & \scriptsize{.335/.337} & \scriptsize{.591/.226} &  \scriptsize{2.631} \\
\scriptsize{ce} & \scriptsize{.335/.592} & \scriptsize{.591/.775} & \scriptsize{-}  \\
\scriptsize{de} & \scriptsize{.337/.592} & \scriptsize{.226/.775} &  \scriptsize{-} 
\end{tabular}
\end{table} \ \\

\vspace{-1.10cm} 
The first reversal, for the triad ($a$,$c$,$d$), tuple ($c$,$d$) has the ratio of eigenvector values from Section~\ref{evec} using the third and fourth elements ($c$,$d$), as 0.335/0.337 ($<1$). The corresponding ratio considering the triad $(a,c,d)$ (the fourth sub matrix in Step 1) is 0.845/0.392 ($>1$), indicating a reversal of preferences between the pair $c$ and $d$. The magnitude of this reversal is (0.337/0.335)/(.392/0.845) = 2.168, with larger numbers indicating a larger degree of reversal, and as a consequence, a greater degree of inconsistency. This tuple has another reversal of 2.631 from the triad ($c$,$d$,$e$). 

There are a total of \vspace{-.2cm} 
\begin{itemize}[noitemsep]
\item 7 reversals, viz. 2.168, 2.641, 4.472, 2.477, 2.802, 1.282 and 2.631 
\item the proportion of reversals, with a maximum of 10$\times$3=30 possible reversals, \\$prop3Rev$=7/30=0.2333,
\item the maximum reversal, $max3Rev$=4.472 
\end{itemize}

\subsubsection{Unsupervised Clustering to Generate Consistency Labels}
\label{unsupervisedclustering}

A training dataset of 180,000 Pairwise Comparison Matrices (PCMs) is simulated, with 20,000 PCMs for each matrix dimension (order) ranging from 4 to 12. The simulation method is described in detail in Section~\ref{simulationMethod}.

Two consistency indicators, `prop3Rev', and `max3Rev', are calculated for each PCM as explained in Section~\ref{inconsistencyindicators}. These two indicators are then standardized separately for each order. The standardized indicators are then used to cluster the PCMs, order-wise, into one of two categories - consistent or inconsistent, using k-means clustering. While the binary clustering assigns each PCM a category membership these categories cannot be directly classified as either consistent or inconsistent. To do this, the category with the lower average value of $prop3rev$ (proportion of preference reversals for a PCM) is designated as having the $consistent$ ab-initio classification.

\subsubsection{Developing a Logistic Model for Consistency Classification}
\label{logisticmodel}

We construct a logit classification model based on the identified predictors to determine the outcome class. The outcome class was initially identified through unsupervised k-means clustering.

\begin{knitrout}
\definecolor{shadecolor}{rgb}{0.969, 0.969, 0.969}\color{fgcolor}\begin{kframe}
\begin{verbatim}
## 
## Call:
## glm(formula = as.formula("abinitConsistent ~ order + prop3Rev + max3Rev"), 
##     family = binomial, data = trainData)
## 
## Coefficients:
##               Estimate Std. Error z value Pr(>|z|)    
## (Intercept)    5.07370    0.09172   55.32   <2e-16 ***
## order          4.32041    0.04575   94.43   <2e-16 ***
## prop3Rev    -113.39521    1.19890  -94.58   <2e-16 ***
## max3Rev       -5.22824    0.05593  -93.48   <2e-16 ***
## ---
## Signif. codes:  0 '***' 0.001 '**' 0.01 '*' 0.05 '.' 0.1 ' ' 1
## 
## (Dispersion parameter for binomial family taken to be 1)
## 
##     Null deviance: 171123  on 126008  degrees of freedom
## Residual deviance:  19112  on 126005  degrees of freedom
## AIC: 19120
## 
## Number of Fisher Scoring iterations: 10
\end{verbatim}
\end{kframe}
\end{knitrout}

The log(odds) for a logical PCM being `$consistent$' is
\begin{equation}
\mathrm{z=log(odds)} = 5.07 +4.32 \times \mathrm{order} -113.40 \times \mathrm{prop3Rev}\  -5.23 \times \mathrm{max3Rev}
\end{equation}

Using this model, the probability of the PCM in Section~\ref{inconsistencyindicators} of being consistent, is\\
$\displaystyle \frac{e^z}{1+e^z}$ = \ensuremath{8.63\times 10^{-11}}.

This logit model estimates the likelihood that a given, human-assigned PCM is consistent.
 
The value \ensuremath{8.63\times 10^{-11}} indicates that it is almost certain that the PCM in Section~\ref{inconsistencyindicators} is inconsistent by the proposed PR method. In fact, with a Consistency Index of 0.33 this PCM is also CR-inconsistent.

\section{Results}
\label{results}

\subsection{Method Testing Framework}

To demonstrate the proposed consistency classification model, the following steps are undertaken.
\begin{itemize}[itemsep=0pt]
\item[1.]  simulate a new batch of logical PCMs
\item[2.]  for each PCM compute the two measures of preference reversal, $prop3Rev$ and $max3Rev$
\item[3.]  perform order-wise clustering of the PCMs using the two indicator measures. This will provide the reference or ab initio classifications
\item[4.]  determine the consistency classification for the simulated PCMs using the logit model developed in Section~\ref{logisticmodel}
\item[5.]  perform a benchmark comparison of 
\begin{itemize}[itemsep=0pt]      
\item[(a)] the proposed classifier with the reference classifier, and  
\item[(b)] the CR classifier with the reference classifier
\end{itemize}
\end{itemize}

\subsubsection{Baseline Calibration}
\label{baselinecalibration}

A batch of logical PCMs is simulated, 20000 for each of orders 4 to 12 as in Section~\ref{simulationMethod}. These PCMs are designed, by their process of simulation, to be similar to those developed by human users. The two measures $prop3Rev$ and $max3Rev$ are computed for each PCM. These two measures are standardized using the order-wise mean and standard deviation from the training set of PCMs. An order-wise k-means clustering 
of the simulated PCMs is carried out based on the standardized values of $prop3Rev$ and $max3Rev$. 
This classifies the PCMs into prima facie consistent and inconsistent categories. The result of this classification is displayed in Table~1.

\begin{table}[H]
\centering
\begin{tabular}{rr}
\hline
Order & Consistent \\
\hline
4 & 72.92 \%  \\
5 & 61.82 \%  \\
6 & 58.99 \%  \\
7 & 57.99 \%  \\
8 & 56.90 \%  \\
9 & 55.80 \%  \\
10 & 54.84 \%  \\
11 & 53.31 \%  \\
12 & 51.57 \%  \\
\hline
Overall & 58.24 \% \\
\hline
\end{tabular}
\label{Tab:abinit}
\vspace{1mm}
\caption{Orderwise ab-initio classification of Logical PCMs}
\end{table}
\vspace{-0.4cm}

The result shows that natural, logically assigned Pairwise Comparison Matrices have a fairly uniform chance, across orders, of being adjudged consistent by an unsupervised, binary classifier. This finding is based on the proper usage of the appropriate metrics to characterize the consistency of a PCM. 

The chances of a logical, order 4 PCM being consistent are greater - an observation that stands to reason due to the relative simplicity of assigning comparison ratios for only four alternatives. 
This calibration immediately resolves the complexities associated with the convoluted benchmarking process in the CR method, which relies on `random PCMs', thereby enabling a more reliable evaluation of consistency.

In the Section~\ref{discussion}, this classification is carried out in parallel for CR-consistency, and the implications of the heavily skewed proportion of consistent matrices thereof are discussed.

\subsubsection{Empirical Validation of the PR classifier}

The logistic classifier model described in Section~\ref{logisticmodel} is now used in order to derive the consistency classification of each of the test PCMs.

For the context of classification of Pairwise Comparison Matrices `\textit{specificity}' is assumed to be the proportion of consistent PCMs that are classified as consistent, and `\textit{sensitivity}' is the proportion of inconsistent PCMs that are classified as inconsistent. 

\begin{table}[H]
\centering
\begin{tabular}{rr|rrr}
\hline
Order & Consistent & Accuracy & Sensitivity & Specificity  \\
\hline
4 & 0.7292 & 
    0.9521 & 0.8770 & 0.9800 \\
5 & 0.5690 & 
    0.9714 & 1.0000 & 0.9496 \\
6 & 0.5579 & 
    0.9758 & 0.9999 & 0.9568 \\
7 & 0.5484 & 
    0.9815 & 0.9891 & 0.9752 \\
8 & 0.5332 & 
    0.9690 & 0.9573 & 0.9794 \\
9 & 0.5157 & 
    0.9495 & 0.9138 & 0.9830 \\
10 & 0.5484 & 
    0.9815 & 0.9891 & 0.9752 \\
11 & 0.5332 & 
    0.9690 & 0.9573 & 0.9794\\
12 & 0.5157 & 
    0.9495 & 0.9138 & 0.9830 \\
\hline
All & 0.5824 & 
    0.9683 & 0.9601 & 0.9742 \\
\hline
\end{tabular}
\vspace{1mm}
\caption{Preference Reversal (PR) Classification Result}
\end{table} 
A uniformly high rate of both sensitivity as well as specificity across matrix orders is a testimony to the effectiveness of the Preference Reversal method for consistency detection. In a batch of 20000 PCMs for each of the orders 4 to 12, the number of inaccurate classifications, False Positives plus False Negatives, is only 3.17\%.

\section{Method Comparison \& Discussion}
\label{discussion}

The ab-initio consistency classification of logical Pairwise Comparison Matrices (PCMs) based on order-wise k-means clustering provides a natural and prima facie evaluation. This methodology is particularly effective because the variables used for k-means clustering serve as natural proxies for consistency.

In Section~\ref{baselinecalibration}, we observed that a logical PCM has a slightly higher probability of being ab-initio consistent, with this probability gradually decreasing as matrix orders increase. However, in real-world scenarios, higher-order matrices would likely exhibit a lower chance of consistency compared to the findings in Table~1. This discrepancy arises because the algorithm used to simulate logical PCMs employs a fixed range for perturbation of the best comparison ratio in the Fundamental Scale. For a human expert, the increase in the number of alternatives exacerbates cognitive fuzziness, resulting in decrease in the precision of judgment.

Table~3 clearly shows that the Preference Reversal (PR) based consistency evaluation, derived from the logit model developed in this paper, closely replicates the unsupervised classification, thereby lending it significant validity. In contrast, the Consistency Ratio (CR) classifier results in a substantially lower proportion of CR-consistent matrices, with this proportion declining sharply as the number of alternatives increases. Overall, only 2.6\% of logical PCMs of orders 7 to 12 are CR-consistent, which serves as a clear empirical refutation of the CR consistency evaluation approach when benchmarked against the unsupervised, ab-initio classification.

\begin{table}[H]
\centering
\begin{tabular}{rccc}
\hline
      & \multicolumn{3}c{PCMs adjudged Consistent using} \\
      & Unsupervised  & Consistency Ratio & Preference Reversal \\
Order & Classification & CR Method & PR Method\\
\hline
4 & 72.92 \%  & 28.12 \%  & 74.80 \%  \\
5 & 61.82 \%  & 20.24 \%  & 64.81 \%  \\
6 & 58.99 \%  & 12.59 \%  & 59.58 \%  \\
7 & 57.99 \%  & 7.39 \%  & 55.97 \%  \\
8 & 56.90 \%  & 4.08 \%  & 54.03 \%  \\
9 & 55.80 \%  & 2.02 \%  & 53.39 \%  \\
10 & 54.84 \%  & 1.13 \%  & 53.97 \%  \\
11 & 53.31 \%  & 0.58 \%  & 54.21 \%  \\
12 & 51.57 \%  & 0.28 \%  & 54.87 \%  \\
\hline
Overall & 58.24 \% & 8.49 \%  & 58.40 \%   \\
\hline
\end{tabular}
\vspace{1mm}
\caption{Orderwise Unsupervised, CR and PR classification of Logical PCMs}
\end{table} 
\vspace{-0.4cm}

To compare the PR method discussed in the previous section with the extant CR classification, the results of the two classification methods for 20,000 `logical PCMs' simulated for each order, are displayed in Table~4 below. Here, as elsewhere in this paper, the true classification is taken to be the orderwise unsupervised classification of a PCM. This provides a basis for evaluating the comparative accuracy of our PR approach versus the CR classification of the PCMs.

\begin{table}[!ht]
\centering
\begin{tabular}{rc|ccc|ccc} 
\hline
& Ab-initio & \multicolumn{3}{c|}{Preference Reversal (PR)} & \multicolumn{3}{c}{Consistency Ratio (CR)} \\ 
Order & Consistent & Consistent & Sensitivity & Specificity & Consistent & Sensitivity & Specificity \\ 
\hline
4 & 0.729 & 
    0.748 & 
    0.877 & 
    0.980 & 
    0.281 &
    0.996 &
    0.384 \\
5 & 0.618 & 
  0.648 & 
    0.907 & 
    0.991 & 
  0.202 & 
    0.996 &
    0.325 \\
6 & 0.590 & 
  0.596 & 
    0.966 & 
    0.986 & 
  0.126 & 
    0.999 &
    0.213 \\
7 & 0.580 & 
  0.560 & 
    0.998 & 
    0.964 & 
  0.074 & 
    1.000 &
    0.127 \\
8 & 0.569 & 
  0.540 & 
    1.000 & 
    0.950 & 
  0.041 & 
    1.000 &
    0.072 \\
9 & 0.558 & 
  0.534 & 
    1.000 & 
    0.957 & 
  0.020 & 
    1.000 &
    0.036 \\
10 & 0.548 & 
  0.540 & 
    0.989 & 
    0.975 & 
  0.011 & 
    1.000 &
    0.021 \\
11 & 0.533 & 
  0.542 & 
    0.957 & 
    0.979 & 
  0.006 & 
    1.000 &
    0.011 \\
12 & 0.516 & 
  0.549 & 
    0.914 & 
    0.983 & 
  0.003 & 
    1.000 &
    0.006 \\
\hline
\end{tabular}
\vspace{1mm}
\caption{Classification of 20,000 Logical PCMs of orders 4 to 12, using PR and CR methods}
\end{table}

The comparison of the PR and CR methods using non-random logical PCMs, simulated to resemble human assigned PCMs, shows that the CR method is highly sensitive to inconsistency. In fact, sensitivity is a perfect 1 for orders 9 and above. On the other hand the specificity is very low, dipping to less than 10\% for orders 8 and above. This inadequacy is corrected by the PR method that has high sensitivity as well as high specificity.
The results also highlight the order bias in the case of CR classification. Specificity, in particular is a casualty, a poor 35\% for orders 4 and 5, the proportion dropping to less than 10\% for orders eight and above.

\subsection{Examples highlighting benefits of the PR approach}

The Preference Reversal method of consistency evaluation is an organic approach based on fundamentals. A reasonable expectation for a consistent matrix is that the preference order between two alternatives would remain invariant over different 3-tuple alternative subsets. While the Fundamental Scale significantly eases the cognitive burden of choosing between multiple available alternatives, it also imposes constraints for the pairwise comparison ratios that make the attainment of CR-consistency, difficult. A more grounded approach is to look for preference reversals - the proportion of reversals as well as the maximum reversal - as possible indicators of consistency. 

The CR consistency approach places far too large a premium on consistency that, for higher orders, becomes practically impossible. Logical PCMs, simulated as described, are more tolerant of human errors of judgement, or moderate biases and thereby are more representative of naturally assigned comparison matrices. 

The range of values of the two consistency indicators, \textit{prop3Rev} and \textit{max3Rev}, are presented in scatter diagrams in Figure~1 for two sets of 2000 (i) simulated Logical PCMs of orders 6, 8, 10 and 12, and (ii) Random PCMs of the same four orders which have been made CR-consistent by an iterative process due to \citet{Harker1987}\footnote{
A CR-consistent PCM is obtained from a random PCM by repeatedly using Harker's method to modify the largest contributor to inconsistency in a PCM}.

\begin{knitrout}
\definecolor{shadecolor}{rgb}{0.969, 0.969, 0.969}\color{fgcolor}\begin{figure}[H]
\includegraphics[width=\maxwidth]{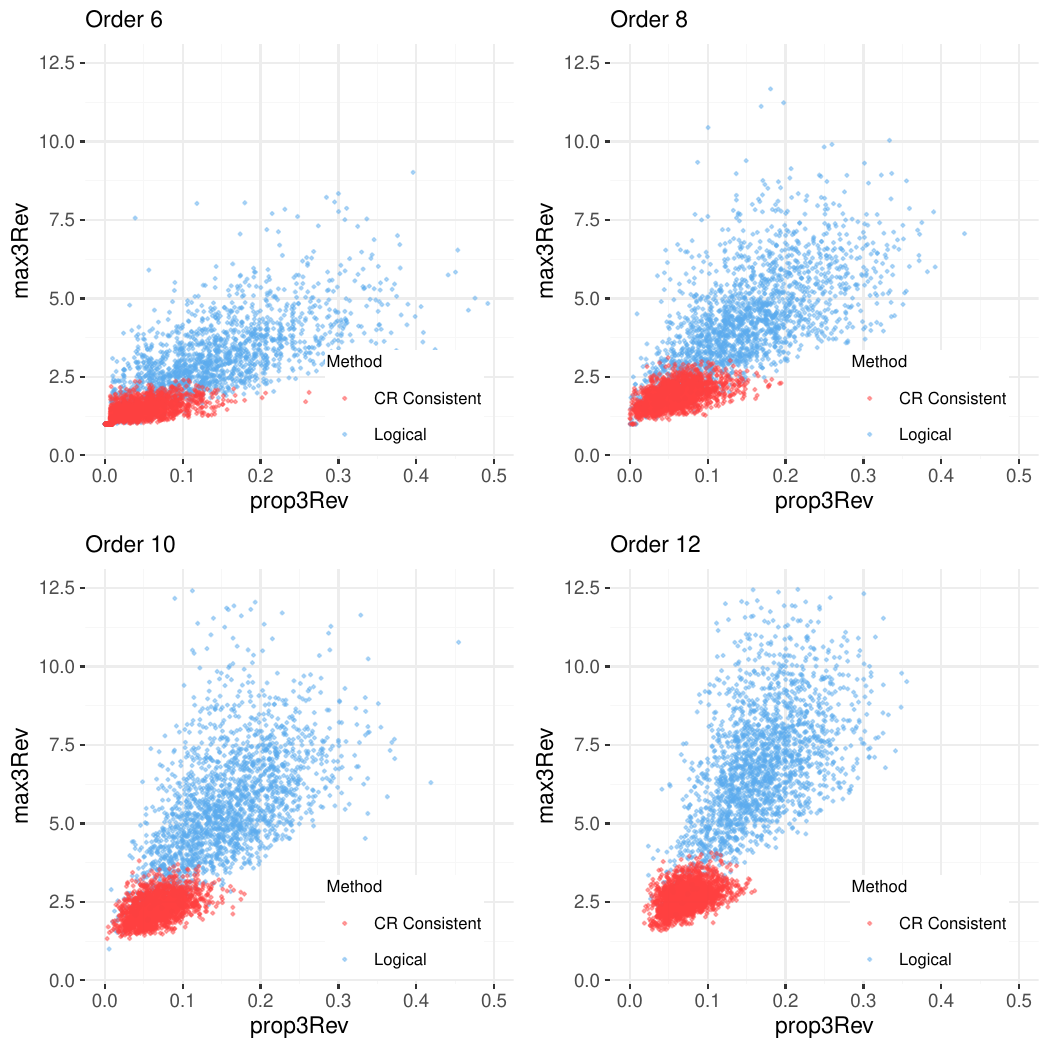} \caption[PR consistency measures for simulated Logical and CR-Consistent PCMs]{PR consistency measures for simulated Logical and CR-Consistent PCMs}\label{fig:ggplot}
\end{figure}

\end{knitrout}
 
The very low values for CR-consistent PCMs for both the consistency indicators can be seen to contrast with the far greater spread of the indicators for Logical PCMs. Viewing this from another perspective, logical PCMs are likely to have a higher number of reversals as well as a larger value of maximum reversal. The corollary to this observation is that a CR consistent matrix is likely to show an unnaturally low level of reversals. This raises the concern of false negatives - PCMs wrongly classified as inconsistent - in consistency evaluation using the CR approach, while also making the task of CR-consistency achievement difficult for naturally assigned pairwise comparison matrices.

For a simulation of 20000 logical PCMs for each of orders 4 to 12, 50\% of CR inconsistent PCMs were consistent using the PR classification. For a batch of 20000 random PCMs which have been coerced to CR-consistency by using \citet{Harker1987}, 639 PCMs of order 4 and 639 PCMs of order 9 are PR inconsistent, demonstrating a small proportion of false positives using the CR-classification approach. More importantly it is possible that the process of coercing CR consistency could well alter preferences indicated in the original PCM.

Out of 20,000 simulated logical PCMs of order 6, 8085, or 40.42\% are PR inconsistent.

Two examples of PCMs where the PR and the CR method differ in their classification of consistency are discussed here. 

\subsubsection{Example of a PCM that is CR-consistent but is PR-inconsistent}
\begin{table}[!ht]
\centering
\begin{tabular}{>{\centering\arraybackslash}m{1.0cm} l|rrrrrr|}
& & \small{\text{a}} & \small{\text{b}} & \small{\text{c}} & \small{\text{d}} & \small{\text{e}} & \small{\text{f}}\\
\cline{2-8}
\multirow{6}{*}{$A$ = }   & \small{\text{a}} & \small{1} & \small{1} & \small{1} & \small{1/2} & \small{1/2} & \small{1}  \\
 & \small{\text{b}} & \small{1} & \small{1} & \small{1/3} & \small{1/2} & \small{1/2} & \small{1} \\
 & \small{\text{c}} &\small{1} & \small{3} & \small{1} & \small{1/2} & \small{1/2} & \small{1/2} \\
 & \small{\text{d}} &\small{2} & \small{2} & \small{2} & \small{1} & \small{1} & \small{7} \\
 & \small{\text{e}} &\small{2} & \small{2} & \small{2} & \small{1} & \small{1} & \small{5} \\
 & \small{\text{f}} &\small{1} & \small{1} & \small{2} & \small{1/7} & \small{1/5}  & \small{1} \\
\cline{2-8}
\end{tabular}
\end{table}

This PCM, $A$ has an eigenvalue of 6.5636 and is CR-consistent. The principal eigenvector of $A$ is (0.241, 0.212, 0.288, 0.651, 0.583, 0.223).

There are 13 preference reversals, with a maximum reversal 2.0556 considering all the $\tbinom{6}{3}$ triad subsets formed from the six alternatives. Three of these reversals occur in the triad submatrix for ($a$,$c$,$f$). The principal eigenvector (0.558, 0.442, 0.702) along with the reversals is shown here.

\begin{table}[H]
\centering
\begin{tabular}{l|rrr| >{\centering\arraybackslash}m{1cm}  >{\centering\arraybackslash}m{.4cm}|>{\centering\arraybackslash}m{1.5cm}>{\centering\arraybackslash}m{1.5cm}>{\centering\arraybackslash}m{.75cm}}
& \small{\text{a}} & \small{\text{c}} & \small{\text{f}} & & \small{abc} & \small{full} & \small{triad} & \small{rev} \\
\cline{1-4} \cline{6-9}
\small{\text{a}} & \small{1} & \small{1} & \small{1} & &  \small{ac} & \small{.241/.288} & \small{.116/.923} &  \small{1.508} \\
\small{\text{c}} &\small{1} & \small{1} & \small{1/2} & & \small{af} & \small{.241/.223} & \small{.116/.366} & \small{1.363} \\
\small{\text{f}} &\small{1} & \small{2} & \small{1} & & \small{cf} & \small{.288/.223} & \small{.923/.366} &  \small{2.056} \\
\cline{1-4} \cline{6-9}
\end{tabular}
\end{table}

The maximum reversal for this PCM occurs for the elements $c$, $f$ in the triad ($a$,$c$,$f$) and is $max3Rev$=2.0513. The preference ratio for c:f from the entire 6 element eigenvector is .288/.223 = 1.2914 whereas the corresponding ratio for the tuple ($c$,$f$) is .442/.702 = 0.6296. The corresponding reversal in preferences for `$c$' and `$f$' is 1.2361/.6296 = 2.0513. The 13 preference reversals, out of a maximum $\tbinom{6}{3} \times 3$, works out to a proportion of $prop3Rev$=0.2167. 

Using Eqn.(1) from Section~\ref{consistencyModel} the probability of this PCM being consistent is only 
0.0131, thus classifying PCM A as PR-inconsistent.

\subsubsection{Example of a PCM that is CR-inconsistent but is PR-consistent}
\begin{table}[!ht]
\centering
\begin{tabular}{>{\centering\arraybackslash}m{1.0cm} l|rrrrrr|}
& & \small{\text{a}} & \small{\text{b}} & \small{\text{c}} & \small{\text{d}} & \small{\text{e}} & \small{\text{f}}\\
\cline{2-8}
\multirow{6}{*}{$B$ = }   & \small{\text{a}} & \small{1} & \small{9} & \small{6} & \small{1/3} & \small{2} & \small{1/2}  \\
 & \small{\text{b}} & \small{1/9} & \small{1} & \small{1/5} & \small{1/9} & \small{1/9} & \small{1/9} \\
 & \small{\text{c}} &\small{1/6} & \small{5} & \small{1} & \small{1/2} & \small{1/3} & \small{1/5} \\
 & \small{\text{d}} &\small{3} & \small{9} & \small{2} & \small{1} & \small{1/2} & \small{1/2} \\
 & \small{\text{e}} &\small{1/2} & \small{9} & \small{3} & \small{2} & \small{1} & \small{1/2} \\
 & \small{\text{f}} &\small{2} & \small{9} & \small{5} & \small{2} & \small{2}  & \small{1} \\
\cline{2-8}
\end{tabular}
\end{table}

This PCM, $B$ has an eigenvalue of 6.6252 and is CR-inconsistent. The principal eigenvector of $B$ is (0.443, 0.043, 0.138, 0.448, 0.399, 0.651). \\

There are 4 preference reversals, with a maximum reversal 2.038 
considering all the $\tbinom{6}{3}$ triad subsets formed from the six alternatives. The triad submatrices with reversals are shown here. \\

\begin{table}[H]
\centering
\begin{tabular}{l|rrr| >{\centering\arraybackslash}m{1cm}  >{\centering\arraybackslash}m{.4cm}|>{\centering\arraybackslash}m{1.5cm}>{\centering\arraybackslash}m{1.5cm}>{\centering\arraybackslash}m{.75cm}}
& \small{\text{a}} & \small{\text{d}} & \small{\text{e}} & & \small{ade} & \small{full} & \small{triad} & \small{rev} \\
\cline{1-4} \cline{6-9}
\small{\text{a}} & \small{1} & \small{1/3} & \small{2} & &  \small{ad} & \small{.443/.448} & \small{.498/.653} &  \small{-} \\
\small{\text{d}} &\small{3} & \small{1} & \small{1/2} & & \small{ae} & \small{.443/.399} & \small{.498/.570} & \small{1.270} \\
\small{\text{e}} &\small{1/2} & \small{2} & \small{1} & & \small{de} & \small{.448/.399} & \small{.653/.570} &  \small{-} \\
\cline{1-4} \cline{6-9}
\end{tabular}
\end{table}

\vspace{-0.25cm}
\begin{table}[H]
\centering
\begin{tabular}{l|rrr| >{\centering\arraybackslash}m{1cm}  >{\centering\arraybackslash}m{.4cm}|>{\centering\arraybackslash}m{1.5cm}>{\centering\arraybackslash}m{1.5cm}>{\centering\arraybackslash}m{.75cm}}
& \small{\text{b}} & \small{\text{d}} & \small{\text{e}} & & \small{bde} & \small{full} & \small{triad} & \small{rev} \\
\cline{1-4} \cline{6-9}
\small{\text{b}} & \small{1} & \small{1/9} & \small{1/9} & &  \small{bd} & \small{.043/.448} & \small{.074/.532} &  \small{-} \\
\small{\text{d}} &\small{9} & \small{1} & \small{1/2} & & \small{be} & \small{.043/.399} & \small{.074/.844} & \small{-} \\
\small{\text{e}} &\small{9} & \small{2} & \small{1} & & \small{de} & \small{.448/.399} & \small{.532/.844} &  \small{1.780} \\
\cline{1-4} \cline{6-9}
\end{tabular}
\end{table}

\vspace{-0.25cm}
\begin{table}[H]
\centering
\begin{tabular}{l|rrr| >{\centering\arraybackslash}m{1cm}  >{\centering\arraybackslash}m{.4cm}|>{\centering\arraybackslash}m{1.5cm}>{\centering\arraybackslash}m{1.5cm}>{\centering\arraybackslash}m{.75cm}}
& \small{\text{c}} & \small{\text{d}} & \small{\text{e}} & & \small{cde} & \small{full} & \small{triad} & \small{rev} \\
\cline{1-4} \cline{6-9}
\small{\text{c}} & \small{1} & \small{1/2} & \small{1/3} & &  \small{cd} & \small{.138/.448} & \small{.256/.466} &  \small{-} \\
\small{\text{d}} &\small{2} & \small{1} & \small{1/2} & & \small{ce} & \small{.138/.399} & \small{.256/.847} & \small{-} \\
\small{\text{e}} &\small{3} & \small{2} & \small{1} & & \small{de} & \small{.448/.399} & \small{.466/.848} &  \small{2.043} \\
\cline{1-4} \cline{6-9}
\end{tabular}
\end{table}

\vspace{-0.25cm}
\begin{table}[H]
\centering
\begin{tabular}{l|rrr| >{\centering\arraybackslash}m{1cm}  >{\centering\arraybackslash}m{.4cm}|>{\centering\arraybackslash}m{1.5cm}>{\centering\arraybackslash}m{1.5cm}>{\centering\arraybackslash}m{.75cm}}
& \small{\text{d}} & \small{\text{e}} & \small{\text{f}} & & \small{def} & \small{full} & \small{triad} & \small{rev} \\
\cline{1-4} \cline{6-9}
\small{\text{d}} & \small{1} & \small{1/2} & \small{1/3} & &  \small{de} & \small{.448/.399} & \small{.318/.505} &  \small{1.780} \\
\small{\text{e}} &\small{2} & \small{1} & \small{1/2} & & \small{df} & \small{.448/.651} & \small{.318/.802} & \small{-} \\
\small{\text{f}} &\small{3} & \small{2} & \small{1} & & \small{ef} & \small{.399/.651} & \small{.505/.802} &  \small{-} \\
\cline{1-4} \cline{6-9}
\end{tabular}
\end{table}

The maximum reversal for this PCM occurs for the elements $d$, $e$ in the triad ($c$,$d$,$e$). The preference ratio for d:e from the entire 6 element eigenvector is .448/.399 = 1.1228 whereas the corresponding ratio for the tuple ($d$,$e$) is .466/.848 = 0.5495. The corresponding reversal in preferences for `$d$' and `$e$' is 1.1228/.5495 = 2.0433. The 4 preference reversals, out of a maximum $\tbinom{6}{3} \times 3$, works out to a proportion of 0.0667. 

The consistency measures for this PCM are thus $prop3Rev$=0.0667 and $max3Rev$=2.0433. Using Eqn.(1) from Section~\ref{consistencyModel} the probability of this PCM being consistent is 0.999997. It can be concluded that the PCM B is PR-consistent.

\subsubsection{Differences in conistency evaluation exemplified in PCMs A and B}

Using CR method, PCM A (eigenvalue 6.5636) is consistent while B (eigenvalue 6.6252) is inconsistent. Yet PCM A has a much larger proportion of preference reversals compared to B, making A PR-inconsistent, while B is PR-consistent. 

The differences in consistency evaluation using the extant CR proposed PR approach for PCMs A and B are illustrated in Figure~2 and summarized in Table~5. \\
\begin{table}[H]
\centering
\begin{tabular}{c|ccl|cccl}
\hline
PCM & eigen value & CR & CR consistency & prop3Rev & max3Rev & logit & PR consistency \\
\hline
A & 6.5636 & 0.09 & \textit{CR consistent} & 0.2167 & 2.0513 & 0.0131 & \textit{PR inconsistent} \\
B & 6.6252 & 1.10 & \textit{CR inconsistent} & 0.0667 & 2.0433 & 1.0000 & \textit{PR consistent} \\
\hline
& & & & & & & 
\end{tabular}
\caption{CR and PR consistency for PCMs A \& B}
\end{table}
\vspace{-0.25cm}
The slightly higher CR value makes B CR inconsistent, while A is CR-consistent. Using the PR method, A has 13 preference reversals against only 4 for PCM B. This makes A PR-inconsistent while B is PR-consistent.

\begin{knitrout}
\definecolor{shadecolor}{rgb}{0.969, 0.969, 0.969}\color{fgcolor}\begin{figure}[H]

{\centering \includegraphics[width=\maxwidth]{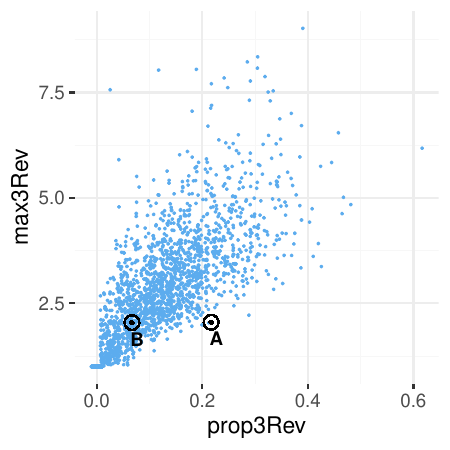} 

}

\caption[PR consistency measures for Logical PCMs (order 6)]{PR consistency measures for Logical PCMs (order 6): A and B superimposed}\label{fig:ggplot2}
\end{figure}

\end{knitrout}
Figure~2 shows that the proposed organic consistency measures clearly differentiate the PCMs A and B, with A having a high number of reversals indicating inconsistency.

\section{Conclusion}

This paper introduces a novel method for assessing the consistency of Pairwise Comparison Matrices (PCMs) in the Analytical Hierarchy Process (AHP) using triadic preference reversals and machine learning techniques. This approach addresses the limitations of the traditional Consistency Ratio (CR) method by providing a more accurate and robust consistency classification. The use of unsupervised k-means clustering and binary logistic regression for detecting consistency achieved a remarkable 97\% accuracy with 2.6\% false negatives, significantly outperforming the CR method's 50\% accuracy and a staggering 85.5\% false negatives.

The results of the simulations demonstrate the efficacy of using logical PCMs, generated by perturbing perfectly consistent matrices within a range derived from the Fundamental Scale, as opposed to random PCMs. This approach not only provides a more realistic benchmark for consistency but also enhances the reliability of the consistency classification process.

The ability to accurately identify inconsistencies has significant implications for decision-making processes in various fields that utilize AHP. By ensuring a more reliable consistency assessment, decision-makers can make more informed and objective choices, ultimately improving the outcomes of multi-criteria decision analysis.

Future research should focus on empirical validation of this method using real-world PCMs from diverse application domains to further demonstrate its practical applicability. Additionally, exploring the integration of this approach with other consistency improvement strategies and comparing its performance with other state-of-the-art methods will provide a deeper understanding of its strengths and limitations.

In summary, the proposed method represents a significant advancement in consistency assessment for AHP. It offers a promising alternative to traditional methods, paving the way for more accurate and reliable decision-making frameworks.






\section*{Declarations}
\subsection*{Disclosure of Interest}
I have no conflicts of interest to disclose.

\subsection*{Declaration of Funding}
No funding was received for this paper.

\nocite{}
\bibliographystyle{apalike}
\bibliography{Consistency3.9} 

\end{document}